\documentclass[letter,traditabstract]{aa} 
\usepackage{graphicx}
\usepackage{lscape}
\usepackage{verbatim}
\usepackage{multirow}
\usepackage{lscape}
\usepackage{txfonts}
\usepackage{booktabs}
\usepackage{bigstrut}
\usepackage{natbib}
\bibpunct{(}{)}{;}{a}{}{,}

\newcommand{\hh}{H$_2$}
\newcommand{\co}{CO(3$-$2)}
\newcommand{\hhs}{H$_2$~1$-$0~S(1)}
\begin{document}
   \title{
ALMA CO and VLT/SINFONI H$_2$ observations of the Antennae overlap region:
mass and energy dissipation
\thanks{Based on ALMA Science Verification data
and observations with the VLT/SINFONI, Program IDs 383.B-0789 and
386.B-0942.}}

   \author{C. N. Herrera
          \inst{1}\thanks{supported by a CNRS-CONICYT grant.},
          F. Boulanger
          \inst{1},
          N. P. H. Nesvadba
          \inst{1},          
         \and
         E. Falgarone
         \inst{2}
          } 
   \institute{Institut d'Astrophysique Spatiale, UMR 8617 CNRS, Universit\'e Paris-Sud 11,
             91405 Cedex Orsay, France 
             \and
            LERMA, UMR 8112 CNRS, Ecole Normale Sup\'erieure and observatoire de Paris, Paris, France           }

\abstract
{

We present an analysis of super-giant molecular complexes (SGMCs)
in the overlap region of the Antennae galaxy merger, based on ALMA
CO(3$-$2) interferometry and VLT/SINFONI imaging spectroscopy of \hhs\
at angular resolutions of 0.9\arcsec and 0.7\arcsec, respectively. All
but one SGMC have multiple velocity components offset from each other by
up to 150~km~s$^{-1}$. \hh\ line emission is found in all
SGMCs and the kinematics of \hh\ and CO are well matched. \hh/CO line
 ratios vary by up to a factor of 10 among  SGMCs and different velocity
components of the same SGMCs. We also identify the CO counterpart of a
bright, compact source of near-IR \hh\ line emission, which shows no
Br$\gamma$, and was first identified with SINFONI. This source has
the highest \hh/CO line ratio, and coincides with the steepest CO velocity
gradient of the entire overlap region. With a size of 50~pc and a
virial mass of a few $10^7~M_{\odot}$ it is perhaps a pre-cluster
cloud that has not yet formed significant numbers of massive stars.
We present observational evidence that the H$_2$ emission is powered by
shocks, and demonstrate how the \hhs\ and the \co\ lines can be used as
tracers of energy dissipation and gas mass, respectively. The variations
in the H$_2$/CO line ratio may indicate that the SGMCs are
dissipating their turbulent kinetic energy at different rates. The
compact source could represent a short ($\sim 1 $ Myr) evolutionary
stage in the early formation of super-star clusters.

}
\titlerunning{CO and  \hh\ observations of the Antennae overlap region}
\maketitle
%
%
%

\section{Introduction}

Major gas-rich mergers are important sites of star formation and
galaxy evolution in the Universe.  The
Antennae galaxy merger (NGC4038/4039) is an ideal target for studying in
detail how galaxy interactions affect the interstellar medium and star
formation. Most stars in the Antennae form in super-star clusters
(SSCs) with stellar masses up to a few $10^6\ M_\odot$
\citep{whitmore10} located where the two galaxies permeate each other,
the 'overlap region'.  Super-giant molecular complexes (SGMCs)
with masses of several $10^8\ M_\odot$ and sizes of $\sim$500 pc have
been identified in CO(1$-$0) in the overlap region with the OVRO
interferometer \citep{wilson00}. \citet{ueda11} have recently reported
higher resolution ($\sim$ 100 pc), CO(3-2) observations of the
Antennae obtained with the SMA.

The formation of SSCs involves a
complex interplay of merger-driven gas dynamics, turbulence fed by the
galaxy interaction, and dissipation of the kinetic energy of the gas.
Hydrodynamic simulations suggest that massive complexes of  cold
gas, akin to SGMCs, form where gas flows trigger compression, cooling
and gravitational fragmentation \citep{teyssier10}.  Within SGMCs, a
hierarchy of structures must form including clouds that are sufficiently massive
to form SSCs \citep{weidner10}.

Recent VLT/SINFONI imaging spectroscopy of the peak of pure-rotational H$_2$ 
emission in the overlap region previously observed with Spitzer \citep[][H11
hereafter]{herrera11} revealed bright diffuse H$_2$ line emission  associated with an SGMC
and a compact ($\sim$0.6\arcsec, $\sim$50 pc) source.  H11 proposed
that the compact source may be a massive cloud on its way to form a
SSC within the next few Myr. The \hh\ lines are powered by shocks and
trace energy that is being dissipated and radiated away as the cloud
complex, and a pre-cluster cloud (PCC) within, grow through gas
accretion.
 
\citet{herrera11} had sub-arcsecond resolution SINFONI data of shocked gas, but lacked CO observations at similar resolution, which are required to probe the bulk of the gas on the relevant scales of $\le$100~pc. In this letter, we take
advantage of the recently  released ALMA science verification 
observations of \co\ in band~7 ($\sim$345~GHz) at $\sim$1\arcsec\
angular resolution to compare the morphology and kinematics of
\co\ and \hh\ line emission in the Antennae overlap
region. We show that together the ALMA and VLT observations provide the
complementary (mass and energetics) data needed to characterize the
dynamical state of SGMCs and to search for pre-cluster clouds. 

%
   
\begin{figure*}[ht]
\centering
\includegraphics[width=15.cm]{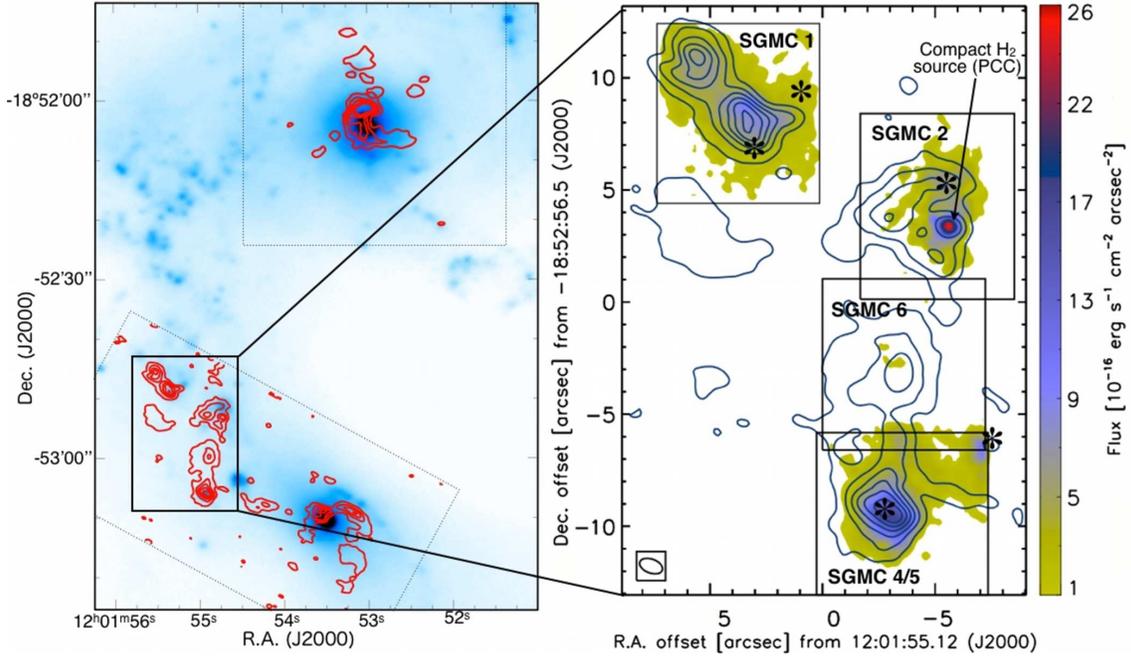}
\caption{{\it Left.}  ALMA CO morphology shown on top of
our CFHT K-band continuum image (H11).  Dotted boxes mark the two
ALMA mosaics, the solid box marks the overlap region. {\it Right.}
\hhs\ morphology as seen with SINFONI. Boxes mark individual SINFONI
fields-of-view, contours show \co\ from 2 to 42 Jy 
km s$^{-1}$ beam$^{-1}$ in steps of 8 Jy 
km s$^{-1}$ beam$^{-1}$. The inset at the bottom left of the
right panel shows the ALMA beam. We also mark massive and young SSCs
(asterisks), and the compact \hh\ source PCC discussed in
\S~\ref{sec:cs}.}
\label{fig:flx}
\end{figure*}

\section{Comparison of ALMA and SINFONI observations}\label{sec:en}

Our analysis relies on two data sets. First,  \co\ line emission was obtained during ALMA science
verification. These data are part of a mosaic of the Antennae
obtained in 10 hr of observing time in band~7 (345~GHz) between
May and June 2011, with 10 to 13 antennae and baselines from 25 to  200~m.
 This gives an angular resolution of
0\farcs6$\times$1\farcs1 (66~pc$\times$115~pc, at a distance of 22~Mpc) and covers the
entire overlap region.  The data have an intrinsic spectral
resolution of 0.85 km s$^{-1}$ and were binned into channels of 10 km
s$^{-1}$. We used the reduced data cubes publicly available on the
ALMA website\footnote{http://almascience.eso.org/alma-data/science-verification},
 which we corrected for the primary beam attenuation. Low spatial
frequencies were filtered out because of missing short spacings that cause a
loss of extended structures $>$4\arcsec and negative sidelobes
adjacent to bright emission. To measure fluxes we used a clipped cube
where all pixel values $<$2$\sigma$ (6~mJy beam$^{-1}$) were set to zero.  To
quantify the missing flux, we constructed spectra at the center of the
overlap region with the angular resolution of the single-dish JCMT and
HHT CO(3$-$2) observations of \citet{zhu03} and \citet{schulz07}, respectively. 
We found about half the total flux of the single-dish
data. The total flux of the SGMCs in the overlap region agrees with
the SMA observations \citep{ueda11}.
Second, we used VLT/SINFONI imaging spectroscopy of
ro-vibrational \hh\ lines in the near-infrared $K$-band at R$=$3000 in four regions
each 8\arcsec$\times$8\arcsec\ in size (Fig.~\ref{fig:flx}). They
were observed in February 2011 with on-source integration times of 40
minutes per pointing. Our previous observations of SGMC~2 have been discussed
by H11. We obtained and reduced the data of the three additional fields in a similar way.

\begin{figure*}[ht]
\centering
\includegraphics[width=18.4cm]{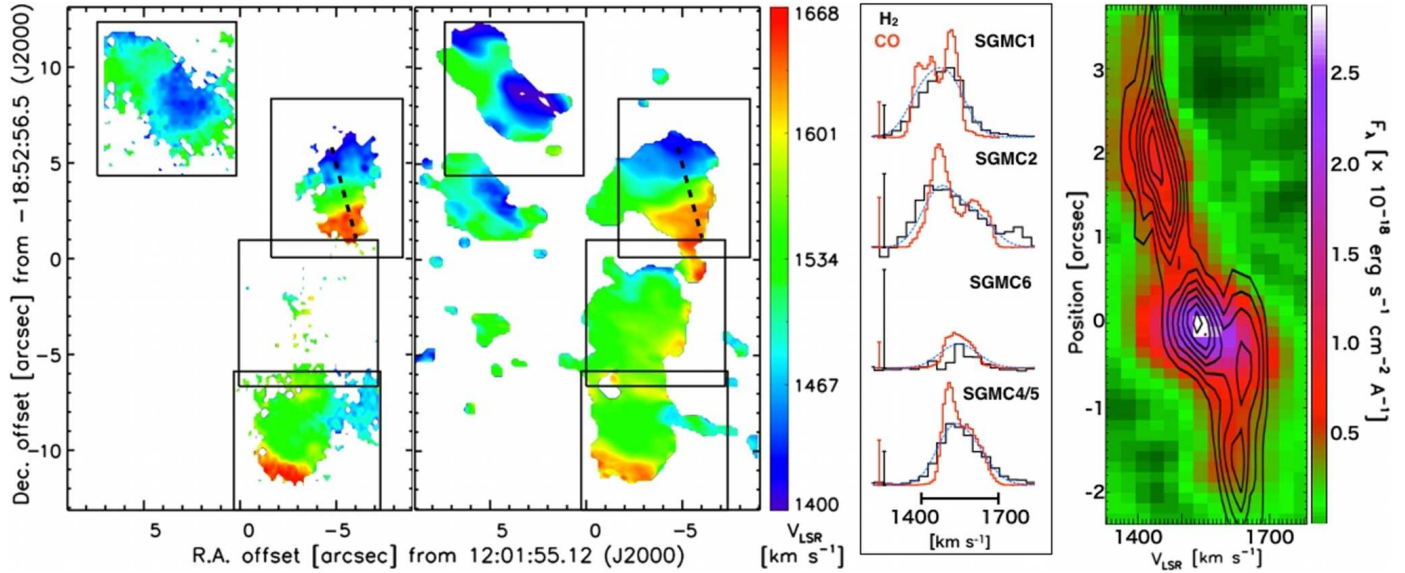}
\caption{{\it Left panel:} Velocity map of \hhs\ (left) and first-moment map of \co\ (right). Dotted lines mark the position-velocity cut shown in the right panel.
{\it Mid panel:} Integrated line profiles of \co\ (red) and \hhs\ (black) for each SGMC. Blue spectra show the \co\ lines convolved to the spectral resolution of SINFONI. Black bars correspond to 5 $\times 10^{-16}$ erg s$^{-1}$ cm$^{-2}$ $\AA^{-1}$  for \hhs\, and red bars to 1.5 Jy  for \co, for all SGMCs. Offsets between the black and blue spectra indicate 
variations in the CO-to-\hh\ ratio between velocity components.  {\it Right panel:}
\hhs\ position-velocity diagram of SGMC~2 (H11), with \co\ emission shown as contours in steps of 0.02 Jy beam$^{-1}$ starting at 0.03 Jy beam$^{-1}$.}\label{fig:vel}
\end{figure*}

Fig.~\ref{fig:flx} shows the spatial distribution of the \hh\ and CO
emission in the Antennae overlap region at a spatial resolution of
$\le$1\arcsec\ (100~pc). The right panel displays the color image of
the \hhs\ emission with \co\ contours for comparison.  Intensity maps
were obtained from the SINFONI and ALMA data cubes by fitting Gaussian
profiles to the spectra in each spatial pixel, using one component for
SINFONI and up to three components for the higher spectral resolution
ALMA data. In Fig.~\ref{fig:vel} we compare the velocity fields of
these two lines. For the CO velocities we computed the first moment map, and
for \hh\ we constructed the velocity map from Gaussian fits.

The ALMA channel maps are very similar to the SMA observations in \citet{ueda11}. We
identified all SGMCs discovered by \citet{wilson00} except for SGMC~3, which
is not covered by ALMA. The new data show two velocity components in SGMC~1
and 2. We also decomposed SGMC~4/5 into two smaller complexes at positions
($-$3,$-$9.5) and ($-$3,$-$3.5) in Fig.~\ref{fig:flx}, each of which has two velocity
components. We kept the name SGMC4/5 for the first complex, and labeled
the  northern extension SGMC~6. CO and \hh\ spectra of
each SGMC are displayed in Fig.~\ref{fig:vel}. The CO spectra were integrated over
each box in Fig.~\ref{fig:flx} using the clipped cube.

Table~\ref{tab:comp} lists the CO line properties of each SGMC and
each velocity component, named a and b. Fluxes, velocities and line
widths of all components are measured from Gaussian fits to 
the spectra in Fig.~\ref{fig:vel}. Error bars include only fit uncertainties, not
the systematic errors owing to the missing short spacings. We estimated
the R$_{3-2/1-0}=I_{\rm CO(3-2)}/I_{\rm CO(1-0)}$ ratios of the SGMCs
by comparing the ALMA and OVRO data, after smoothing to a common
resolution. $I_{\rm CO}$ is the integrated intensity in
K~km~s$^{-1}$. This ratio varies from source to source between
0.3$-$0.8, with a mean value of 0.5, consistent with what
\citet{schulz07} found from single-dish data for the entire overlap
region, as well as the peak line ratios measured with the SMA for each
individual SGMC \citep{ueda11}.

We estimated gas masses from the CO fluxes, where the $X_{\rm
CO}$ factor is the main source of uncertainty. To be consistent with
previous studies, we used the same $X_{\rm CO}$ factor for
CO(1$-$0) as \citet{wilson00}, $X_{\rm CO}$$=$$3\times
10^{20}$~H$_2$~cm$^{-2}$~(K~km~s$^{-1})^{-1}$, and adopted the scaling
R$_{3-2/1-0}$$=$0.59 of \citet{schulz07}.
Our mass estimates (Tab.~\ref{tab:comp})
are comparable to those of \citet[][]{wilson00}. 
Similar to H11, our data show that most of the H$_2$ emission
away from SSCs is powered by shocks, not UV heating in PDRs. The
observed H$_2$~2$-$1/1$-$0~S(1) ratios, 0.1$-$0.2 (Tab.~\ref{tab:comp}) can be
accounted for by PDR models, but only for high UV fields ($\chi$$>$$10^4$, in units of the mean value in the solar neighborhood) and high densities \citep[$n_{\rm H}$$>$$10^5$
cm$^{-3}$,][]{lepetit06}. These conditions exist near massive stars embedded in molecular clouds, but the extended H$_2$ emission is generally not observed to peak near the brightest SSCs. The notable exception is the embedded
cluster in SGMC4/5, which, however, has a higher line ratio
\citep{gilbert00}. In addition, the [CII]157$\mu$m/[OI]63$\mu$m line
ratio indicates that the mean radiation field in the overlap region is
$\chi$$\sim$$10^3$ \citep{schulz07}, an order of magnitude smaller than that
required to account for the H$_2$~2$-$1/1$-$0~S(1) ratios. The other luminous SSCs have already dispersed most of their gas and dust. Thus, the brightest UV-heated gas would come from clouds away from the clusters. Outside the clusters the intensity of the radiation field is not very high. For a 5$\times$10$^6~M_{\odot}$ SSC, the UV field is $\chi$$>$$10^4$ only out to a
distance of $<$100 pc from the cluster and less if we include extinction.

Our analysis has three main results. (1) All SGMCs have \hhs\
line emission and the \hhs\ kinematics match those of \co\ 
well. \citet{zhu03} and \citet{schulz07} found that the CO is emitted
from gas at temperatures with T$\sim$30$-$150~K. The excitation temperature of the warmer H$_2$ gas emitting in the near-IR is $\sim$1000$-$2000~K (H11). The
similarity of the gas kinematics of CO and \hh\ indicates that warm and
cold gas are closely associated. (2) Figs.~\ref{fig:flx} and
\ref{fig:vel} show large variations in the \hh/CO ratio between SGMCs
and between individual velocity components in the same SGMC
(Tab.~\ref{tab:comp}).  
Extinction is an unlikely cause, because the near- and mid-IR
H$_2$ pure-rotational emission-line regions have similar morphologies \citep[at least at the 5\arcsec\ scales resolved by Spitzer-IRS; see
Fig.~4 in ][and the discussion in H11]{brandl09}. In H11, we related the \hh\ emission to the
dissipation of kinetic energy. 
With this interpretation, the \hh/CO ratio 
traces differences in the energy dissipation rate per unit mass, which
must be related to the dynamical evolution of the gas. (3) 
All clouds have two spatially separated velocity components as seen in
the channel maps of SGMC~2 in Fig.~\ref{fig:shock}. In the other clouds, the
spatial offsets between velocity components are less obvious, possibly
because of projection effects. The velocity difference between components
within an individual SGMC is up to 150~km~s$^{-1}$ (Fig.~\ref{fig:vel}). Given the size 
and mass of the SGMCs, this is too 
large to be accounted for by the gas self-gravity. The gas kinematics is 
most likely driven by the galaxy interaction. Single-dish observations found 
similar velocity gradients in the
extended emission around the SGMCs, which further supports this idea. 
The two components in SGMC~2 
are spatially separated and resolved by ALMA. We estimated their sizes and found their virial masses 
to be comparable to the molecular masses derived from the CO fluxes.
\vspace{-3mm}

\begin{table*}[ht]
\begin{center}
\caption{\co\ properties of the SGMCs (top) and PCC (bottom). We include the integrated 
\hhs\ fluxes, and the \hh~1$-$0/2$-$1~S(1) and \hh/CO flux ratios. For PCC, the line parameters are measured with a single-component Gaussian fit.}\label{tab:comp}
\begin{tabular}{ccccccccc}
\hline \hline
\noalign{\smallskip}
   
\multirow{2}{*}{Source} & Velocity &  $V_{\rm LSR}$ &  $\Delta$v  &$S_{\rm CO}$ & $M_{\rm mol}$  & $F_{\rm H_2~1-0~S(1)}$& \multirow{2}{*}{$ F_{\rm H_2~2-1~S(1)} \over F_{\rm H_2~1-0~S(1)} $} &  \multirow{2}{*}{$\rm H_2~1-0~S(1) \over CO(3-2) $ }\\   
\noalign{\smallskip}
             &  component &km s$^{-1}$   & km s$^{-1}$ & Jy km s$^{-1}$ &$M_{\odot}$ & erg s$^{-1}$ cm$^{-2}$ & &\\
\hline
\noalign{\smallskip}
\multirow{2}{*}{SGMC~1}	&	a &	 1417$\pm$2	& 105$\pm$6	& 380$\pm$24	& 5.6$\pm0.3 \times 10^8$	& \multirow{2}{*}{1.6$\times 10^{-14}$}	& \multirow{2}{*}{0.2}	& \multirow{2}{*}{2.1}	 \\
					&	b &	1521$\pm$1	& 63$\pm$3	& 289$\pm$16	& 4.2$\pm0.2 \times 10^8$	&	&	& \\
\multirow{2}{*}{SGMC~2}	& 	a &	1469$\pm$1	& 90$\pm$2	& 289$\pm$8	& 4.2$\pm0.1 \times 10^8$	& \multirow{2}{*}{8.3$\times 10^{-15}$}	& \multirow{2}{*}{0.2}	& \multirow{2}{*}{1.6}	\\
					& 	b & 1613$\pm$2	& 114$\pm$5	& 176$\pm$10	& 2.6$\pm0.1 \times 10^8$	& 	&	& \\
\multirow{2}{*}{SGMC~4/5}& 	a & 1505$\pm$1	& 58$\pm$2	& 171$\pm$9	& 2.5$\pm0.1 \times 10^8$	&  \multirow{2}{*}{1.3$\times 10^{-14}$}	& \multirow{2}{*}{0.3}	&\multirow{2}{*}{2.6}	 \\
					& 	b & 1586$\pm$3	& 126$\pm$6	& 256$\pm$12	& 3.7$\pm0.2 \times 10^8$	& 	&	& \\
\multirow{2}{*}{SGMC~6}	& 	a & 1506$\pm$2	& 50$\pm$8	& 48$\pm$12	& 0.7$\pm0.2 \times 10^8$	& \multirow{2}{*}{6.7$\times 10^{-16}$}	& \multirow{2}{*}{$<$0.5$^{a}$}	&\multirow{2}{*}{0.2}	 \\
					& 	b & 1561$\pm$5	& 113$\pm$6	& 216$\pm$15	& 3.2$\pm0.2 \times 10^8$	& 	&	& \\
\hline
\noalign{\smallskip}
H$_2$ source (PCC)	&$-$ &	1534$\pm$2	&  93$\pm$4	& 7.6	$\pm$0.4	& 1.0$\pm0.1 \times 10^7$ &  7.6$\times 10^{-16}$ &  0.1& {8.7}	 \\
\hline
\end{tabular}
\end{center}

$^{a}$ H$_{2}$~2$-$1~S(1) flux corresponds to an upper limit estimated from the noise.
\end{table*}

%

\section{The compact H$_2$ source}\label{sec:cs}
The ALMA maps also give new insight into the nature of the bright,
compact \hh\ emitter associated with SGMC~2, PCC, recently discovered
by H11. PCC is the brightest \hh\ line emitter in the overlap
region. It is not detected in the
lower-resolution CO(1$-$0) data, but is an emission peak in the SMA and ALMA maps (Fig.~\ref{fig:flx}). 
Isolating the PCC CO counterpart from the surrounding extended emission is difficult with an algorithm like 
CLUMPFIND because of the large velocity gradient across PCC (see right panel in Fig.~\ref{fig:vel}).
This may explain why the source is not specifically listed in Table B.1 of \citet{ueda11}.
There is no other CO peak in the SGMCs with an obvious
counterpart in the near-IR, except for the embedded SSC in SGMC~4/5.

\begin{figure}[ht]
\centering
\includegraphics[width=0.45\textwidth]{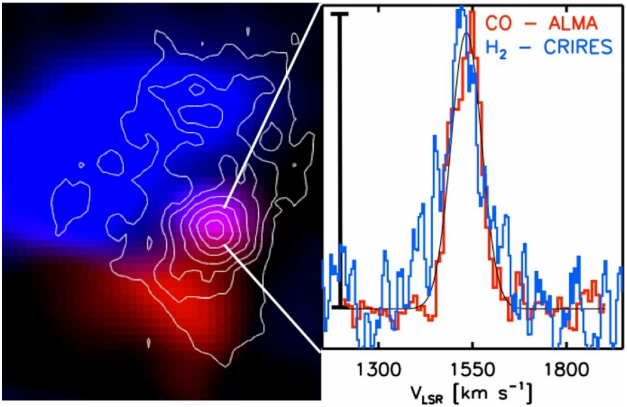}
\caption{({\it left}) \co\ emission from SGMC~2, where emission
between 1350$-$1540 km~s$^{-1}$ and between 1540$-$1750 km~s$^{-1}$
are shown in blue and red, respectively. Contours show the \hhs\
morphology. ({\it right}) \co\ (red) and CRIRES \hh\ (blue) spectrum
of PCC.  The \hh\ spectrum is smoothed to 18 km s$^{-1}$
resolution. The bar corresponds to 77 mJy~beam$^{-1}$ for \co\ and 4.4
$\times 10^{-17}$ erg~s$^{-1}$~cm$^{-1}$~\AA$^{-1}$ for
\hhs.}\label{fig:shock}
\end{figure}

Fig.~\ref{fig:shock} shows a comparison of the CO and CRIRES high-resolution (6 km s$^{-1}$) \hhs\ 
spectra of PCC (H11). 
The CO spectrum is the peak emission at the position of the PCC corrected  for the surrounding
emission, measured over an annulus outside the
source, to isolate the CO velocity component associated with the
PCC. \co\ and \hhs\ spectra are remarkably similar. Table 1 lists the parameters of our Gaussian fit.
The \hh-luminous PCC has a velocity dispersion (40 km~s$^{-1}$)
significantly higher than those of GMCs with the same sizes in the Milky Way \citep[5 km
s$^{-1}$;][]{falgarone09,heyer09}.

Using the velocity dispersion of \co, we
obtain a virial mass of $M_{\rm vir}=5 R \sigma^2/G=4.6\times
10^{7}\times (\sigma_{v}/40~{\rm km~s}^{-1})^2~M_{\odot}$. Since ALMA
does not spatially resolve the PCC, we instead used the 50~pc size measured
with SINFONI. 
 The exceptionally high H$_2$~1$-$0 S(1)-to-Br$\gamma$ line ratio ($>$15, H11) provides unambiguous evidence that the H$_2$ emission of the PCC is powered by shocks \citep{puxley90}. 
The \hh/CO ratio, i.e. the energy dissipation rate per unit mass, is also exceptionally high, a factor 5 higher than that of the SGMC~2 complex overall.

PCC appears to be located at the interface between blue and redshifted
gas (Fig.~\ref{fig:shock}) where CO shows a steep velocity gradient
($\sim$1 km~s$^{-1}$~pc$^{-1}$ in the position-velocity diagram in Fig.~\ref{fig:vel}). The
observed properties of PCC are consistent with a scenario where 
the formation of SSCs is triggered by interactions between two gas flows. 
In SGMC~2, depending on the full three-dimensional
geometry, the flows could either be colliding or creating a large velocity
shear, and most likely a combination of both. In either case, the interaction drives a
turbulent energy cascade in which kinetic energy is being dissipated. 
 This is where we would expect the highest energy dissipation rate.

The bolometric luminosity of the PCC is $\sim$$10^7 \,
L_\odot$. Observations of, e.g., NGC~1333 \citep{maret09} show that
the bolometric luminosity of protostellar outflows is on the order of $10^{5}~
L_{\odot} \times \dot{M}_{\rm wind}$, where $\dot{M}_{\rm wind}$ is the stellar
mass loss rate in M$_{\odot}$ yr$^{-1}$. The small embedded stellar mass of $M_{*}$$=$$4\times 10^4
\, M_\odot$ (H11) makes protostellar winds an implausible energy source.

The cloud luminosity may be accounted for by the dissipation
of the cloud kinetic energy for a cloud mass of a few $10^7 \,
M_\odot$ -- a value comparable to the virial mass $5 \times 10^7 \,
M_\odot$ -- and a dissipation timescale of
1~Myr. This is comparable to the cloud crossing time, and also the
dynamical time scale associated with the velocity gradient of $\rm 1
\, km \, s^{-1} \, pc^{-1}$ at the position of the cloud.  The
similarity of both timescales indicates that the cloud may still be
forming by accreting gas, and therefore that a significant part of the
cloud luminosity may be powered by gas accretion.  In any case, the
time during which the PCC is a bright \hh\ emitter is short, about
1~Myr. 

This short timescale may explain why we do not find more bright compact
sources in our new \hh\ data.
\citet{whitmore10} list five massive ($>$10$^5$ $M_{\odot}$), young ($<$
5~Myr) SSCs over the part of the overlap region covered with both ALMA and SINFONI (Fig.~\ref{fig:flx}). This is
consistent with finding only a single bright PCC with the SINFONI data if the
\hh-luminous phase does not last longer than a few Myr.

Our analysis gives a  foretaste of the power of combining
mass and energy tracers to study the dynamical state of molecular gas
in galaxy mergers and the early stages of the formation of SSCs. In
the future, this approach can be extended with ALMA and VLT using
additional tracers of energy dissipation and mass.

%

\begin{acknowledgements}
\vspace{-1mm}
We wish to thank the staff at ALMA and the VLT for making these
observations and are particularly grateful to the ALMA SV team for
making the fully reduced and calibrated ALMA data available 
(ADS/JAO.ALMA\#2011.0.00003.SV). We thank R. Kneissl for helping us in analyzing the ALMA SV maps, C. Wilson for providing us with her
CO(1$-$0) OVRO data cube, and P. Guillard for his useful comments. We
thank the referee for comments that  improved our paper.
\vspace{-5mm}
\\
\end{acknowledgements}
\bibliographystyle{aa}
\bibliography{papers}
\end{document}